\documentclass[final,3p,times,twocolumn]{elsarticle}
\usepackage{amssymb}
\usepackage{color}

\usepackage[nodots]{numcompress}
\usepackage[none]{hyphenat}
\journal{Current Opinion in Solid State $\&$ Materials Science}

\begin{document}

\begin{frontmatter}



\title{Pairing insights in iron-based superconductors from scanning tunneling microscopy}


\author{Can-Li Song}
\author{Jennifer E. Hoffman}
\cortext[cor1]{Corresponding author: jhoffman@physics.harvard.edu}
%\ead{jhoffman@physics.harvard.edu}
\address{Department of Physics, Harvard University, Cambridge, MA 02138, U.~S.~A.}

\begin{abstract}
Scanning tunneling microscopy (STM) has made tremendous progress in the study and understanding of both classical and unconventional superconductors. This has motivated a rapidly growing effort to apply the same techniques to the iron-based high-$\textit{T}_c$ superconductors since their discovery in 2008. Five years have brought exciting advances in imaging and spectroscopic investigation of this new class of materials. In this review, we focus on several recent STM contributions to the identification of the gap symmetry and pairing glue. We highlight the unique capabilities and challenges still ahead for STM studies of iron-based superconductors.
\end{abstract}

\end{frontmatter}

% \linenumbers

\section{\label{sec:introduction}Introduction}
Magnetism has long been thought to be antagonistic to superconductivity. Therefore, the 2008 discovery of superconductivity in iron-containing LaO$_{1-x}$F$_{x}$FeAs \citep{kamihara2008iron}, with transition temperature $T_{c}$ rapidly climbing to 55 K upon replacement of La by magnetic rare earth elements \citep{ren2008superconductivity}, was unexpected and provoked worldwide excitement. As in the case of cuprates, the superconductivity emerges from the antiferromagnetic parent compounds, suggesting a link between spin fluctuations and electron pairing in both materials \cite{Scalapino2012thread}. This finding opens a new avenue to address the two ultimate goals in the field of superconductivity: to seek the microscopic origin of the high $T_{c}$ and then design new materials with higher $T_{c}$. Despite extensive experimental and theoretical explorations, iron-based superconductors (Fe-SCs) still face fierce debates on a number of issues including gap symmetry, a prerequisite for understanding the secret of high-$T_{c}$ superconductivity. An excellent review by Johnston has thoroughly discussed the puzzle in Fe-SCs through 2010 \citep{johnston2010puzzle}. Another comprehensive review by Stewart covers the rapidly-moving field through 2011 \citep{stewart2011superconductivity}.

Scanning tunneling microscopy (STM) provides unique capabilities to image the atomic and electronic structure of a surface with a sub-unit-cell spatial resolution.  STM has been applied with great success to the study of conventional and cuprate superconductors \citep{fischer2007scanning}. The local density of states directly measured by STM spectroscopy provides indispensable information about the superconducting gap structure, its spatial inhomogeneity and behaviors near impurity as well as vortex core states. Furthermore, via quasiparticle interference (QPI) imaging, STM can provide momentum-resolved information about pairing symmetry, collective excitations, and competing phases. All these accomplishments motivate the growing use of STM to study Fe-SCs in the quest to understand the intricate electron pairing in these materials.

Following an early review of STM studies of Fe-SCs by Yin \textit{\textit{et al}} \citep{yin2009scanning}, STM has made considerable new progress and greatly contributed to the study of some of the most unusual and remarkable properties of these materials. A more thorough review by Hoffman recently discussed the cleaved surface configurations, superconducting and other spectral gaps, and vortex states \citep{hoffman2011spectroscopic}. Our new review here will begin with a brief introduction to the STM technique in section 2, then concentrate on the experimental highlights in the pairing symmetry of Fe-SCs obtained over the past year by STM, such as tunneling spectroscopy in section 3, QPI in section 4, and vortex state in section 5. In section 6, we conclude by mentioning a few very recent hints of higher $T_{c}$ in iron-based materials, as well as suggestions for future STM experiments that should shed additional light on these materials.

\section{\label{sec:STM/STS}Scanning tunneling microscopy and spectroscopy}
STM is based on the quantum tunneling of electrons between two electrodes separated by a thin potential barrier. A sharp metallic tip, which acts as a local probe, is brought within a short distance (typically several \AA) of an electrically conducting sample surface. The tip can be positioned with sub-$\AA$ precision in both the $xy$ plane and the $z$ direction using a three-dimensional piezoelectric scanner, as schematically illustrated in Fig.\ 1(a). Applying a bias voltage between the metallic tip and conducting sample leads to a measurable tunneling current; the polarity of the bias voltage determines the direction of the net electron flow. For instance, a negative bias voltage applied to the sample will allow electrons to tunnel from the occupied states of the sample through the vacuum barrier into the empty states of the tip [Fig.\ 1(b)]. Upon reversing the bias polarity, the electrons will tunnel in the opposite direction, from occupied states of the tip into empty states of the sample. Based on the Tersoff-Hamann theory \cite{Tersoff1983theory}, the tunneling current $I$ can be well approximated by

\begin{equation}
I \propto e^{-2\kappa d},\; \kappa=\frac{\sqrt{2m\phi}}{\hbar}\approx0.5\sqrt{\phi}\,\mathrm{{\AA}}^{-1}
\end{equation}

\noindent where $\phi$ is a mixture of the work functions of the tip and sample measured in e\textit{V} and $d$ is the tip-sample separation measured in \AA. For typical metals, $\phi\sim5$ eV, so $I$ will increase by about one order of magnitude for every $\AA$ decrease in $d$.

\begin{figure}[tbh]
\centering
\includegraphics[width=\columnwidth]{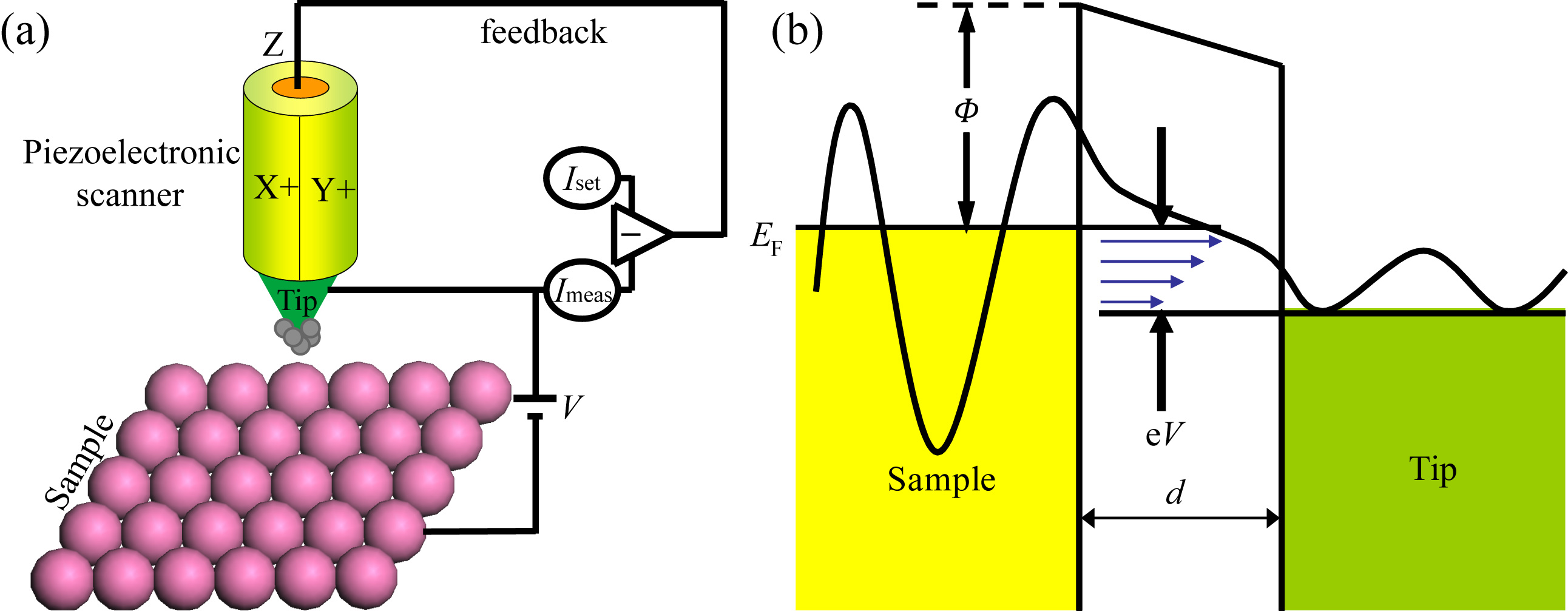}
\caption{(color online) (a) Schematic of an STM. A voltage $V$ is applied between a sharp metallic tip and a conducting sample surface, leading to a measurable tunneling current $I_{\mathrm{meas}}$ which decays exponentially with tip-sample separation $d$. In standard constant-current topographic imaging, the difference (or the error signal) between $I_{\mathrm{meas}}$ and the setpoint current $I_{\mathrm{set}}$ is fed back to the $z$ piezo to control the tip height. (b) Tunneling process of electrons between the tip and sample across a vacuum barrier of width $d$ and height $\phi$. The electron wave functions of both the sample and tip decay exponentially into vacuum with a small overlap, allowing electrons to tunnel between them. With a negative bias voltage $V$ applied to the sample, electrons tunnel from the occupied states of the sample to the empty states of the tip.}
\end{figure}

In topographic mode, the surface is mapped based on the decay of tunneling current $I$ with increasing tip-sample separation $d$. With the bias voltage $V$ fixed, the error signal between the measured current $I_{\mathrm{meas}}$ and the setpoint current $I_{\mathrm{set}}$ is fed back to the $z$ piezo to control $d$ as the tip is rastered across the sample surface [Fig.\ 1(a)]. The $z$ trajectory of the tip therefore maps a contour of constant integrated electron density of states (DOS). This technique is referred as constant-current mode. In the case of homogeneous DOS, the contour corresponds to the geometric topography of the surface. However, if the DOS varies spatially, the resulting image contains a mixture of DOS and geometric information. By setting the tunneling voltage $V_{\mathrm{set}}$ far from the energy range of spatially inhomogeneous states, the contribution of the inhomogeneous DOS can be significantly reduced, so that geometry will dominate the topographic image as desired.

In addition to revealing the geometry of a sample surface, STM can also probe the evolution of sample DOS with energy, up to several e$V$ from the Fermi level ($E_{\mathrm{F}}$) in both occupied and unoccupied states. The DOS can be accessed by switching off the feedback circuit to fix $d$, sweeping the bias voltage $V$, and recording the tunneling current $I(V)$. The conductance $dI/dV$ can be obtained either by numerical differentiation of $I(V)$ or by a lock-in amplifier technique. In the latter case, a small modulation is added to the bias voltage $V$, and the tunneling current $I$ is demodulated to yield $dI/dV$. Although the interpretation of $dI/dV$ spectra can be quite complex, in ideal conditions $dI/dV$ is a good measure of the sample DOS. If these $dI/dV$ spectra are recorded on a dense array of locations in real space with well-chosen $V_{\mathrm{set}}$, the spatial variation of the sample DOS can be extracted. This DOS mapping technique has been applied to measure local gap variations and magnetic vortices \citep{fischer2007scanning,yin2009scanning,hoffman2011spectroscopic}.

\section{\label{sec:Low-Temperature Tunneling Spectroscopy}Tunneling spectroscopy}

Like cuprates, all Fe-SCs are layered compounds. Most can be mechanically cleaved to expose an atomically flat and clean $ab$-surface, suitable for characterization by surface-sensitive probes. To date, STM spectroscopy techniques have revealed a wide distribution of reduced gap values $2\overline{\Delta}/k_B T_c$ across materials \citep{hoffman2011spectroscopic}, varying degrees of gap inhomogeneity within materials (inhomogeneous doped $A$Fe$_{2}$As$_{2}$ \citep{yin2009scanningprl,massee2009nanoscale,shan2011observation} \textit{vs}.\ homogeneous LiFeAs and FeSe \citep{hanaguri2012scanning,song2011direct}), multiple superconducting gaps within a single material due to the multi-band electronic structure \citep{shan2011observation}, non-universal pairing gap symmetry \citep{song2011direct,hanaguri2010unconventional}, and symmetry breaking in both superconducting and parent phases \citep{song2011direct,chuang2010nematic,zhou2011quasiparticle}. Here, we extend the discussion of pairing and normal state symmetry based on recent results from higher energy conductance features, atomic impurity imaging, and normal state spectroscopy.

\subsection{\label{sec:Electronbosoncoupling}Electron-boson coupling}
Collective modes, which couple strongly to electrons and may serve as ``glue'' for Cooper pairing, often appear as additional conductance features at energies beyond the superconducting gap. This higher energy structure was first observed and quantitatively analyzed in conventional superconductor Pb-insulator-Pb tunnel junctions, where minimum in $dI^{2}/dV^{2}$ at positive energy $\Omega+2\Delta$ was found to correspond to phonon mode energy \citep{mcmillan1965lead}. However, the exact shape of the $dI^{2}/dV^{2}$ spectrum should be computed from a \textit{k}-space integral involving the pairing gap, electron density of states, and density of states of the phonon (or other collective mode). In unconventional cuprates and Fe-SCs, the asymmetric electron pairing, complex Fermi surface and multi-band coupling complicate the calculation and preclude any universal relationship between the mode energy $\Omega$ and a regular feature of $dI^{2}/dV^{2}$. Therefore, in different materials, $\Omega$ has been inconsistently extracted from several features in $dI/dV$ or $dI^{2}/dV^{2}$.

Tunneling experiments on cuprates showed a similar dip-hump structure in $dI/dV$ at energies above the gap, and several studies \citep{fischer2007scanning,Renner1995vacuum,DeWilde1998Unusual,zasadzinski2001correlation,niestemski2007distinct} specifically matched the spectral features to the energy of a spin resonance mode $\Omega_{r}$ near ($\pi$, $\pi$) observed by neutron scattering \citep{fong1999neutron,wilson2006resonance}. This supports the importance of spin fluctuations in the pairing mechanism of high-$T_{c}$ cuprates \cite{Scalapino2012thread}.

Soon after the discovery of Fe-SCs, the importance of spin fluctuations was suggested in these materials as well, with the proposal that the high-$T_{c}$ superconductivity arises through spin flip quasiparticle excitations between the $\Gamma$-centered hole pockets and $M$-centered electron pockets (so-called $s_{\pm}$ wave pairing) \citep{mazin2008unconventional,kuroki2008unconventional}. In this context, tunneling spectroscopy is again a powerful method to measure the energies of collective modes coupling to electrons, and to compare these energies to spin resonance modes detected by other techniques.

Fasano \textit{et al} first observed a dip-hump feature by spectroscopic STM in nearly optimally doped 1111-type SmFeAsO$_{0.8}$F$_{0.2}$ ($T_{c}$ =45 K) [Fig.\ 2(a)], and argued that this feature was likely a spin resonance mode \citep{fasano2010local}. The spectral asymmetry makes it difficult to observe the superconducting gap edge and dip-hump structure on the negative energy (filled states) side, similar to early cuprate studies \cite{Renner1995vacuum}, which were later clarified with higher resolution spectra showing features at both polarities \cite{DeWilde1998Unusual}. In SmFeAsO$_{0.8}$F$_{0.2}$ on the positive energy side (empty states), the energy of the presumed collective mode, $\Omega=E_{\mathrm{dip}}-\Delta$, ranges from $\sim$2-8 meV, and locally anti-correlates with the presumed pairing strength $\Delta$. This phenomenology is similar to the local anti-correlation of the presumed phonon \citep{lee2006interplay} and spin resonance \citep{niestemski2007distinct} mode energies with gap energies observed in cuprates. However, $\Omega=E_{\mathrm{dip}}-\Delta\sim$ (0.5-2)$k_B T_c$ (and even the larger $E_{\mathrm{hump}}-\Delta\sim$ (2-3)$k_B T_c$) are much lower than the universal neutron resonance energies $\Omega_{r}\sim4.4k_B T_c$ found in Fe-SCs \citep{wang2012close}. It is possible that surface contamination played a role in this study of uncleaved SmFeAsO$_{0.8}$F$_{0.2}$.

\begin{figure}[tbh]
\centering
\includegraphics[width=\columnwidth]{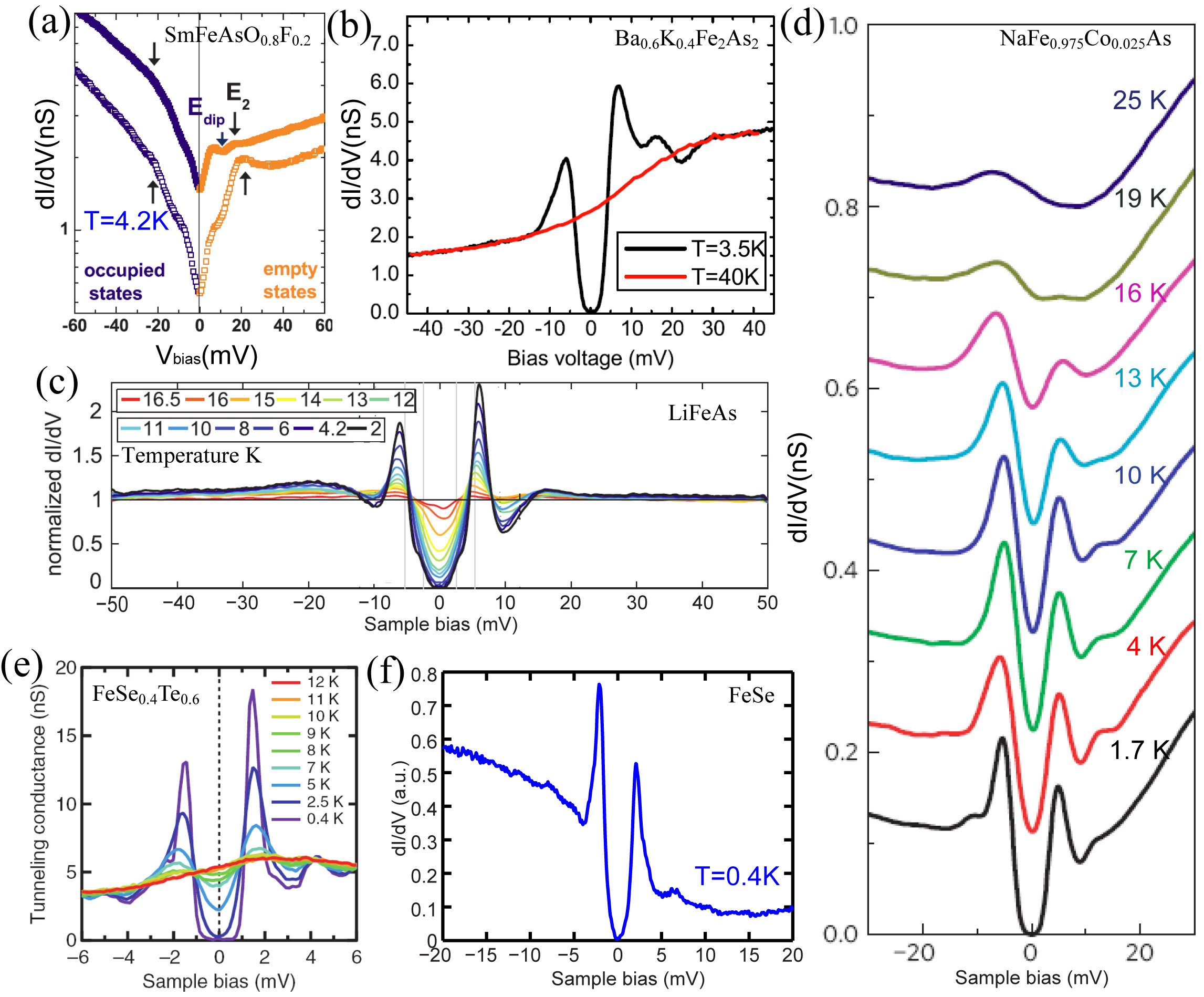}
\caption{Large-energy-scale tunneling conductance spectra showing the dip-hump structure beyond the superconducting gaps in various Fe-SCs: (a) SmFeAsO$_{0.8}$F$_{0.2}$ ($T_{c}$ =45 K) \citep{fasano2010local}; (b) Ba$_{0.6}$K$_{0.4}$Fe$_{2}$As$_{2}$ ($T_{c}$ =38 K) \citep{shan2012evidence}; (c) LiFeAs ($T_{c}$ =17 K) \citep{chi2012scanning}; (d) NaFe$_{0.975}$Co$_{0.025}$As ($T_{c}$ =21 K) \citep{wang2012close}; (e) FeSe$_{0.4}$Te$_{0.6}$ ($T_{c}$ =14.5 K) \citep{hanaguri2010unconventional}; (f) FeSe ($T_{c}$=9.3 K) \citep{song2011direct}.}
\end{figure}

Shan \textit{\textit{et al}} subsequently reported a spectroscopic STM study of hole-doped 122-type Ba$_{0.6}$K$_{0.4}$Fe$_{2}$As$_{2}$ ($T_{c}$ =38 K) \citep{shan2012evidence}, in which features of a collective mode were more clearly visible beyond the superconducting gap in raw $dI/dV$ spectra below $T_{c}$ [Fig.\ 2(b)]. The mode energy of $\Omega\approx$ 13-14 meV, calculated as the difference between the $dI^{2}/dV^{2}$ minimum at the positive voltage and the larger superconducting gap $\Delta\approx$ 8.4 meV, closely agrees with the spin resonance excitation $\Omega_{r}\approx$ 14 meV as measured by neutron scattering \citep{christianson2008unconventional}. Again, $\Omega$ anti-correlates locally with $\Delta$ within this material, which contrasts with the global positive correlation between spin resonance excitation $\Omega_{r}$ and $T_{c}$ across 6 different Fe-SCs \citep{wang2012close}. Balatsky \textit{\textit{et al}} addressed this same discrepancy in cuprates, proposing a local strong coupling model, where $\Delta(r)$ not only scales linearly with the boson mode energy $\Omega(r)$, but also correlates exponentially with the boson coupling constant $g(r)$ via  $\Delta(r)\sim\Omega(r)e^{-1/g(r)}$ \citep{balatsky2006local}. In other words, $g(r)$ is more important than $\Omega(r)$ in determining $\Delta(r)$. In addition, it was found that the effective coupling constant is inversely proportional to the boson energy $\Omega(r)$. Therefore, larger $\Omega(r)$ corresponds to weaker coupling $g(r)$, and consequently to smaller $\Delta(r)$.

To round out the four major families of Fe-SCs, Chi \textit{et al} observed the dip-hump feature in 111-type stoichiometric LiFeAs ($T_{c}$ =17 K) [Fig.\ 2(c)] \citep{chi2012scanning}, and Wang \textit{et al} observed the feature in slightly Co-doped NaFe$_{0.975}$Co$_{0.025}$As ($T_{c}$ =21 K) single crystals [Fig.\ 2(d)] \citep{wang2012close}. Both the deduced bosonic mode energies, $\Omega$ = 5 meV in LiFeAs and $\Omega$ = 7.8 meV in NaFe$_{0.975}$Co$_{0.025}$As, are close to the spin resonance excitation $\Omega_{r}\approx 8$ meV \citep{taylor2011antiferromagnetic}. Moreover, such features are also visible in $dI/dV$ spectroscopy on 11-type iron chalcogenides FeSe$_{0.4}$Te$_{0.6}$ ($T_{c}$ =14.5 K) \citep{hanaguri2010unconventional} and FeSe ($T_{c}$=9.3 K)\citep{song2011direct} [Figs.\ 2(e) and 2(f)], although not clearly explained in these papers.

\begin{figure}[tbh]
\centering
\includegraphics[width=\columnwidth]{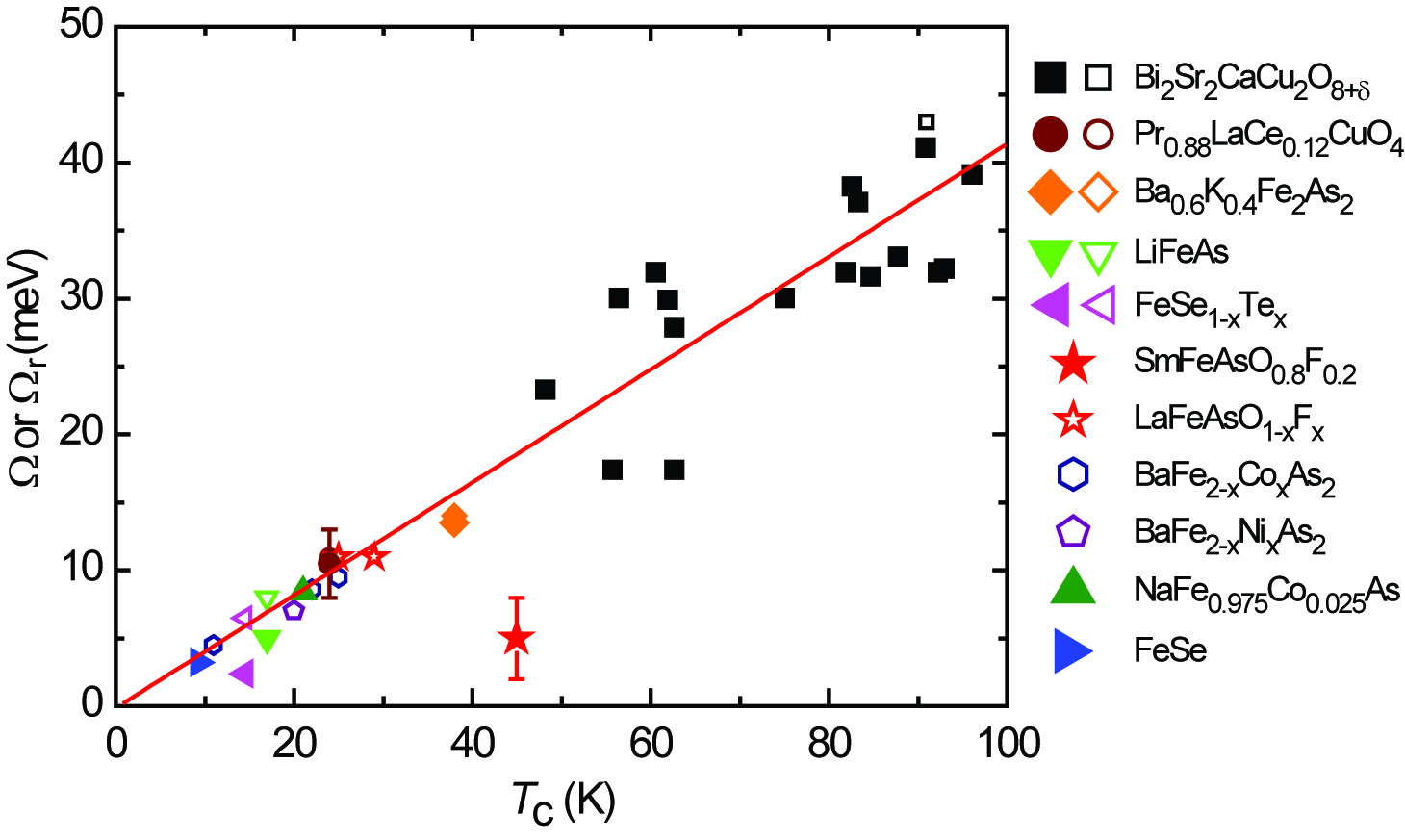}
\caption{Collective resonance mode energy $\Omega$ from STM (filled symbols) and spin resonance $\Omega_r$ from neutron scattering (open symbols) in both cuprates and Fe-SCs \textit{vs}.\ bulk $T_{c}$. The $\Omega$ data for Bi$_2$Sr$_2$CaCu$_2$O$_{8+\delta}$ comes from \textit{dI/dV }spectra from 17 different SIS break junctions at various hole dopings, and is computed as $\Omega = E(\mathrm{dip\ in\ } dI/dV)- 2\Delta$ \citep{zasadzinski2001correlation}. All other $\Omega$ data comes from STM junctions with normal state tips. Pr$_{0.88}$LaCe$_{0.12}$CuO$_4$: $\Omega = E(\mathrm{peak\ in\ } dI^{2}/dV^{2})-\Delta$ \citep{niestemski2007distinct}; SmFeAsO$_{0.8}$F$_{0.2}$: $\Omega = E(\mathrm{dip\ in\ } dI/dV)- \Delta$ \citep{fasano2010local}; Ba$_{0.6}$K$_{0.4}$Fe$_{2}$As$_{2}$: $\Omega = E(\mathrm{dip\ in\ } dI^{2}/dV^{2})-\Delta$ \citep{shan2012evidence}; LiFeAs: $\Omega = E(\mathrm{dip\ in\ } dI/dV)- \Delta$  \citep{chi2012scanning}; NaFe$_{0.975}$Co$_{0.025}$As: $\Omega = E(\mathrm{dip\ in\ } dI^{2}/dV^{2})-\Delta$ \citep{wang2012close}; FeSe$_{0.4}$Te$_{0.6}$ and FeSe: $\Omega = E(\mathrm{peak\ in\ } dI^{2}/dV^{2})-\Delta$ \citep{song2011direct, hanaguri2010unconventional}. Neutron scattering data comes from Refs. \citep{fong1999neutron,wilson2006resonance,wang2012close,christianson2008unconventional,taylor2011antiferromagnetic}.}
\end{figure}

Figure 3 summarizes the bosonic mode energy $\Omega$ derived from STM measurements and $\Omega_r$ from neutron scattering, in both cuprates and Fe-SCs, \textit{vs}.\ the superconducting transition temperature $T_{c}$. Although $T_{c}$ changes by a full order of magnitude, it is remarkable that most data collapse onto the universal relation $\Omega$ (or $\Omega_r)/k_B T_c$ = 4.8 $\pm$ 0.3. Furthermore, $\Omega/2\Delta$ $<$ 1 holds in both cuprates and Fe-SCs \cite{zasadzinski2001correlation,shan2012evidence,chi2012scanning}. These facts suggest that the dip-hump structure observed in both material families may be of the same origin, e.g.\ electron-spin-fluctuation interactions \cite{Scalapino2012thread}. Note that spin fluctuations may extend to high energy in Fe-SCs \cite{dai2012magnetism}. However, the energy dependence of the electron-boson coupling $\alpha^{2}(\omega)$ may be the dominant factor determining the low energy resonance mode $\Omega_{r}$, which leads to a pronounced peak in the pairing glue $\alpha^{2}F(\omega)$ and is visible in the conductance spectra \cite{Ahmadi2011Eliashberg}. In conclusion, the dip-hump structure has been resolved by STM in the four major families of Fe-SCs and in two families of cuprates. These features link to the spin resonance $\Omega_{r}$. The universal relation $\Omega\sim \mathrm{4.8}k_B T_c$ and $\Omega/2\Delta$ $<$ 1 suggests that the interactions of electrons with the spin resonance mode are crucial for the superconductivity in Fe-SCs.

\subsection{\label{sec:Atomicimpurities}Atomic impurities}
Impurities in superconductors are an active subject of study, with the potential to unravel the pairing symmetry of the superconducting state. Impurities are particularly important in cuprates and Fe-SCs, most of which require chemical substitutions to enable superconductivity. While in conventional $s$-wave superconductors, only magnetic impurities cause pair-breaking and suppress superconductivity, in $d$-wave superconductors (e.g.\ cuprates) \citep{balatsky2006impurity} and multi-band $s_{\pm}$ wave superconductors (e.g.\ possibly Fe-SCs) \citep{bang2009impurity,zhang2009nonmagnetic,tsai2009impurity,Masashige2009single,Ng2009ingap,Kariyado2010single,Beaird2012impurity}, both magnetic and non-magnetic impurities are expected to induce bound states in the superconducting gap.

In Fe-SCs, intensive theoretical studies of single impurity effects have been performed since their discovery \citep{bang2009impurity,zhang2009nonmagnetic,tsai2009impurity,Masashige2009single,Ng2009ingap,Kariyado2010single,Beaird2012impurity}. It was proposed that studies of single nonmagnetic impurity can address the ongoing debate between the $s_{\pm}$ and $s_{++}$ pairing symmetry, since most theoretical models suggest that only the sign change of the order parameter in $s_{\pm}$ pairing can give rise to robust sub-gap bound states around non-magnetic impurities \citep{zhang2009nonmagnetic,tsai2009impurity,Ng2009ingap,Kariyado2010single,Beaird2012impurity}. Therefore, spectroscopic study of the energetic structure of impurity-induced bound state resonances could constrain the pairing symmetry. Furthermore, STM study of the spatial structure of impurity-induced states could address proposed phases such as orbital ordering which break the underlying crystal lattice symmetry \citep{lv2009orbital,lee2009ferro}. Exploration of single impurities is most promising in nominally stoichiometric superconductors with unreconstructed surfaces such as FeSe(001), KFe$_{2}$Se$_{2}$(110), and LiFeAs(001) \citep{song2011direct,li2011phase,Grothe2012Bound}.

Single atom impurities were first identified on stoichiometric FeSe films grown by molecular beam epitaxy (MBE) \citep{song2011direct}. Both Fe adatoms and Se vacancies showed sub-gap resonances in $dI/dV$ at symmetric energies, with more pronounced filled state peaks [Figs.\ 4(a)-4(c)]. These sub-gap resonances would indeed be expected at impurity sites for unconventional $s_{\pm}$ wave pairing in FeSe \citep{zhang2009nonmagnetic,tsai2009impurity,Ng2009ingap,Kariyado2010single,Beaird2012impurity}. However, the existence of resonances here does not prove the $s_{\pm}$ pairing, as the magnetic nature of the impurities is unknown: if magnetic, they would possibly induce bound states even in the case of simple $s$-wave pairing \citep{balatsky2006impurity}.

\begin{figure}[tbh]
\centering
\includegraphics[width=\columnwidth]{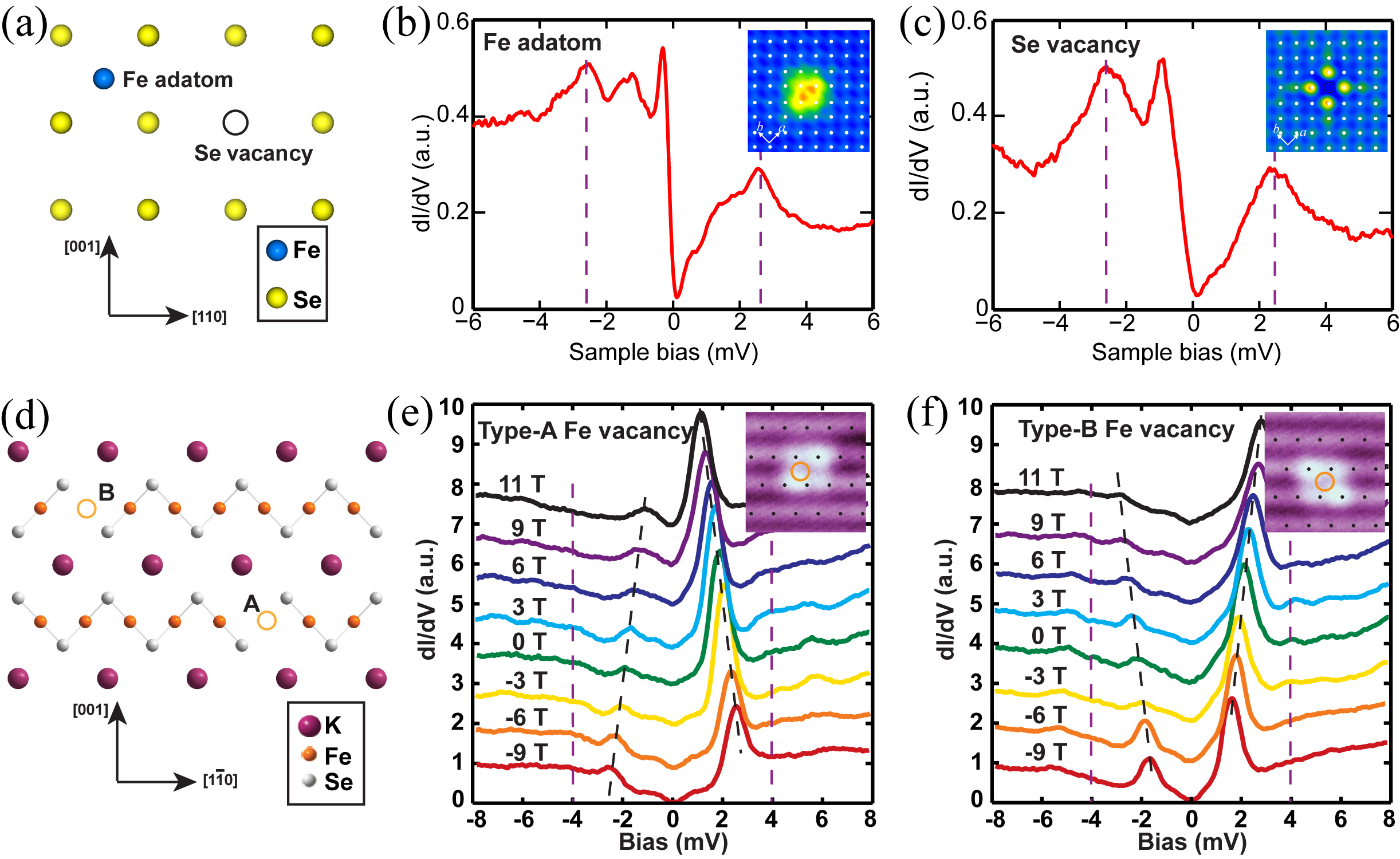}
\caption{Impurity-induced bound states in the superconducting gaps of MBE-grown FeSe(001) and KFe$_{2}$Se$_{2}$(110) films. (a-c) Schematic atomic structure of Fe adatom and Se vacancy, and their effects on the $dI/dV$ spectra at 0.4 K. Insets show the corresponding topographic images (3 nm $\times$ 3 nm, $V_{\mathrm{s}}$ = 10 mV, $I$ = 0.1 nA) with white dots indicating the topmost Se atoms. Pronounced bound states were observed at -1.4 meV and -0.4 meV on Fe adatoms, and at -1.0 meV on Se vacancies. The tunnel junction was set at 10 mV and 0.1 nA \citep{song2011direct}. (d-f) Atomic structure and magnetic-field dependence of $dI/dV$ spectra measured on two inequivalent Fe vacancies at 0.4 K. Insets show the corresponding topographic images (3 nm $\times$ 3 nm, $V_{\mathrm{s}}$ = 30 mV, $I$ = 0.03 nA) with black dots indicating the K atoms. Black dashes depict the evolution of in-gap bound states with the applied magnetic field. The tunnel junction was set at 15 mV and 0.1 nA \citep{li2011phase}. Megenta dashes indicate the energy positions for the superconducting gaps, as determined from spectra acquired at impurity-free regions.}
\end{figure}

Subsequently, Li \textit{et al} introduced two types of mirror-symmetric Fe vacancies [Fig.\ 4(d)] into superconducting KFe$_{2}$Se$_{2}$(110) films \citep{li2011phase}. The $dI/dV$ spectra on both impurities reveal strongly suppressed coherence peaks and a pair of in-gap resonances peaked at $\pm$1.9 mV [Figs.\ 4(e) and 4(f)]. Although the energies of the resonances at both sites are again symmetric with respect to $E_{\mathrm{F}}$, their strengths are again quite different, but with more pronounced empty state peaks in contrast to FeSe \citep{song2011direct}. More interestingly, the energies of the in-gap bound states on the two vacancies were found to shift linearly in opposite directions with applied magnetic field [Figs.\ 4(e) and 4(f)], demonstrating that the two Fe vacancies carry spins of different orientations. Therefore, the observed in-gap resonances do not contradict the theoretical predictions for either $s_{\pm}$ or $s_{++}$ pairing symmetry.

\begin{figure}[tbh]
\centering
\includegraphics[width=\columnwidth]{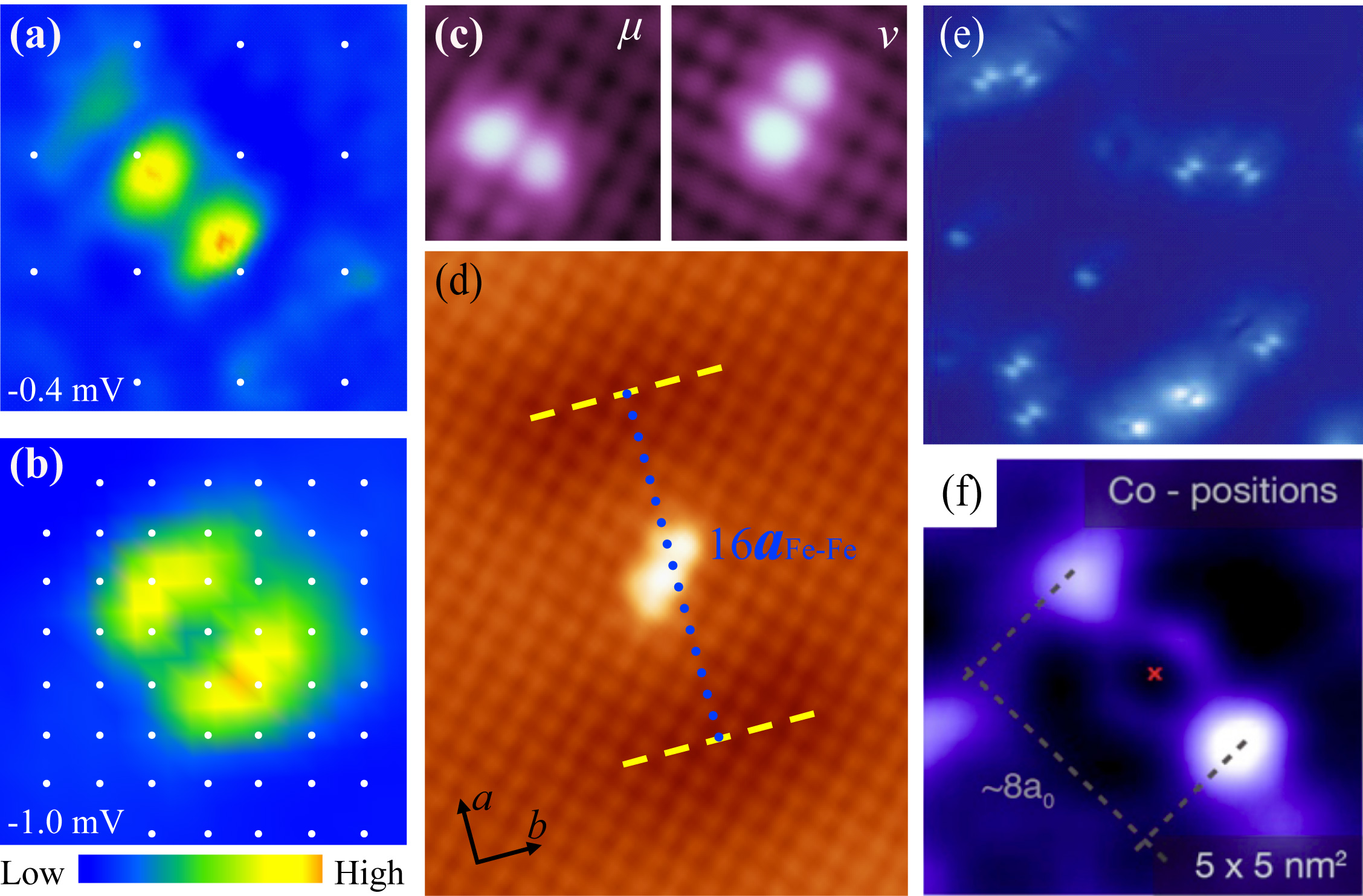}
\caption{Symmetry breaking in Fe-SCs. (a, b) Bound states maps in the vicinity of single Fe adatom (1.5 nm $\times$ 1.5 nm) and Se vacancy (3 nm $\times$ 3 nm) of FeSe films \citep{song2011direct}. The white dots indicate the topmost Se atoms. (c, d) Geometric (2 nm $\times$ 2 nm, $V_{\mathrm{s}}$ = 6 mV, $I$ = 0.1 nA) and electronic (6 nm $\times$ 8 nm, $V_{\mathrm{s}}$ = 10 mV, $I$ = 0.1 nA) dimers induced by the Se substitution for Fe sites in FeSe films \citep{song2012suppression}. (e) Geometric dimers in LiFeAs single crystals (19 nm $\times$ 19 nm, $V_{\mathrm{s}}$ = 20 mV, $I$ = 10 pA) lack accompanying electronic dimers \citep{hanaguri2012scanning}. (f) Average electronic dimer induced by Co substitution at the Fe site in Ca(Fe$_{0.97}$Co$_{0.03}$)$_{2}$As$_{2}$ (5 nm $\times$ 5 nm) \citep{allan2012anisotropic}.}
\end{figure}

Very recently, Grothe \textit{et al} systematically characterized five predominant defects in nominally stoichiometric LiFeAs single crystals \citep{Grothe2012Bound}. Despite their unknown chemical nature, these defects can be named according to their two-dimensional point group symmetries, and categorized into two distinct sets. One set consists of two different Fe-D$_{2}$ defects at the Fe sites, which preserve the local lattice symmetry, and exhibit only a single resonance near the edge of the smaller gap in this multi-gap superconductor. The other set, including Fe-C$_{2}$, As-D$_{1}$ and Li-D$_{1}$, was found to break the local lattice symmetry, and induce additional in-gap bound states, pronounced at either positive or negative biases. All five of these defects suppress the superconducting coherence peaks, and their in-gap resonances show such pronounced particle-hole asymmetry in the spectral weight that for some resonances no corresponding symmetric peak can be detected on the opposite side of $E_{\mathrm{F}}$. Although the magnetic nature of the impurities is unknown, one may speculate that some of the five may be non-magnetic, in which case the consistent existence of sub-gap resonances at all five different impurities would match theoretical predictions for $s_{\pm}$ but not $s_{++}$ pairing symmetry \citep{zhang2009nonmagnetic,tsai2009impurity,Ng2009ingap,Kariyado2010single,Beaird2012impurity}.

In addition to the bound state spectra of these impurities, which provide important information about pairing symmetry, the spatial structure of single atom defects, which often breaks the local lattice symmetry, supports the proposed orbital ordering phase \citep{lv2009orbital,lee2009ferro}. In FeSe, taking into account the full three-dimensional structure of each FeSe sheet, both Fe adatoms and Se vacancies have structural C$_{4v}$ symmetry (aside from the unresolvably small 0.4\% orthorhombicity \citep{song2011direct}), yet both of their bound state $dI/dV$ maps show pronounced C$_{2v}$  symmetry [Figs.\ 5(a) and 3(b)]. This likely originates from an orbital ordering effect \citep{song2011direct}. Furthermore, Se substitutions at the Fe site were also identified \citep{song2012suppression}. Although these defects produce stochastically oriented short-range ``geometric dimer'' signatures in keeping with the local C$_{2v}$ symmetry of their lattice sites  [Fig.\ 5(c)], they are also found to produce longer range $a$-axis-oriented electronic dimers with peaks separated by $\sim16a_{\mathrm{Fe-Fe}}$ [Fig.\ 5(d)].

The two orthogonal Fe-D$_{2}$ ``geometric dimer''  in LiFeAs \citep{Grothe2012Bound} can be tentatively assigned to the two inequivalent Fe positions, by comparison to the analogous Se substitutions at Fe sites in FeSe films \citep{song2012suppression}. Interestingly, the Fe-D$_{2}$ impurities in LiFeAs do not appear to produce electronic dimers [Fig.\ 5(e)] \citep{hanaguri2012scanning}, suggesting that neither orbital ordering \citep{lv2009orbital,lee2009ferro} nor pocket density wave ordering \citep{kang2012dimer}, both of which may give rise to such dimers, arises in the tetragonal phase of Fe-SCs such as LiFeAs.

Finally, although the density of Co dopants substituted at the Fe sites in orthorhombic Ca(Fe$_{0.97}$Co$_{0.03}$)$_{2}$As$_{2}$ is too high to clearly distinguish special feature around individual Co aotms, an atomically registered averaging analysis recently showed that each dopant also forms an $a$-axis-oriented electronic dimer [Fig.\ 5(f)], similar to the electronic dimers in FeSe, but with peaks separated by 22 \AA, approximately $8a_{\mathrm{Fe-Fe}}$ \citep{allan2012anisotropic}.

\subsection{\label{sec:Normalstate}Normal state}
STM spectroscopy was also used to probe the electronic structure of iron-based compounds in their parent and normal states. Most such studies have revealed a quite broad and asymmetric V-shaped DOS suppression around $E_{\mathrm{F}}$ \citep{hoffman2011spectroscopic}, which is reminiscent of the controversial pseudogap in cuprates \citep{fischer2007scanning}. In the case of iron-based materials, however, several other phenomena, such as surface reconstructions, or structural and magnetic transitions above $T_{c}$, may be responsible for the DOS suppression around $E_{\mathrm{F}}$. It therefore appears more challenging to understand the DOS suppression in iron-based materials.

\begin{figure}[tbh]
\centering
\includegraphics[width=\columnwidth]{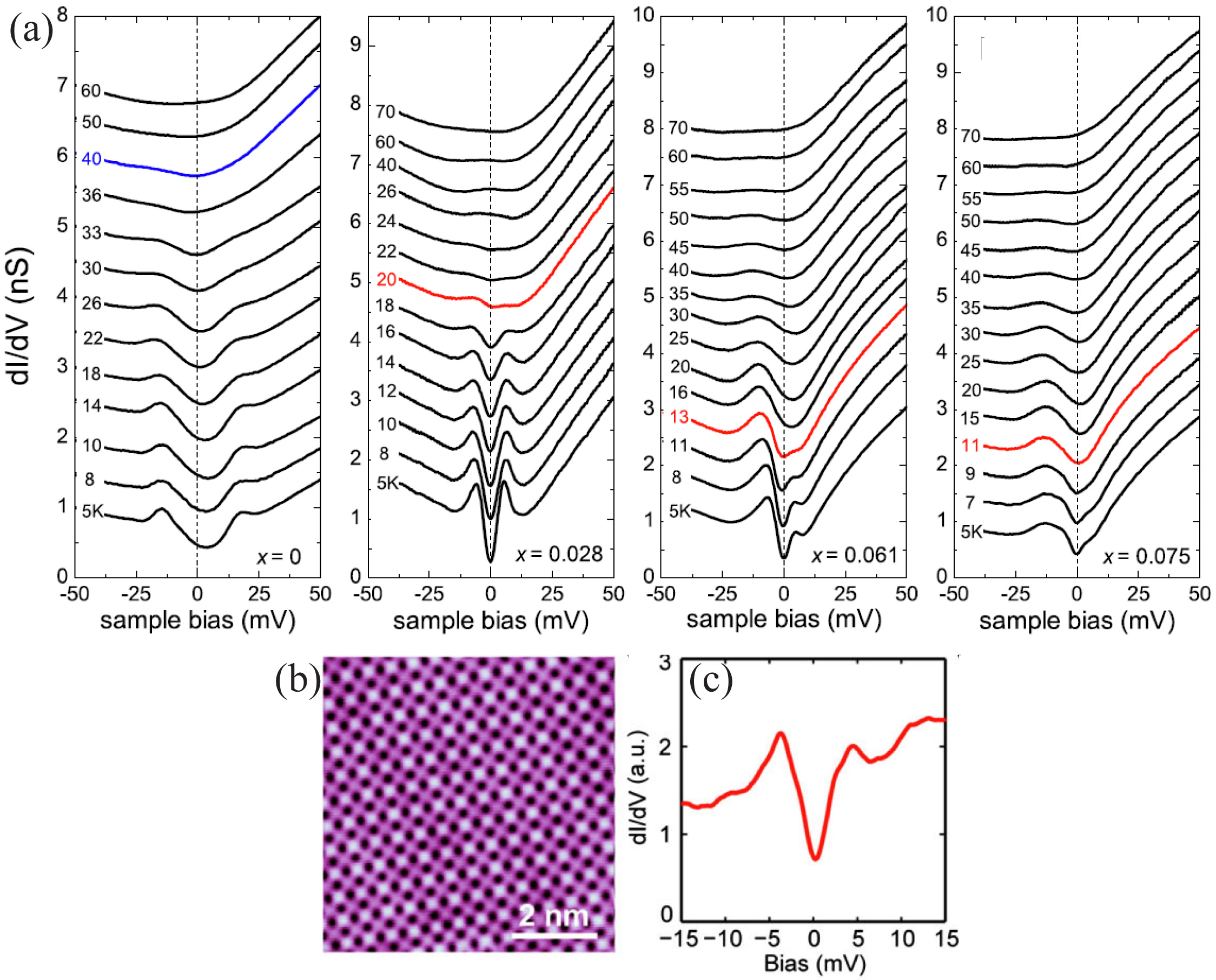}
\caption{(a) Temperature dependence of $dI/dV$ spectra in NaFe$_{1-x}$Co$_{x}$As single crystals with various Co concentrations \textit{x}. The blue (red) curve marks SDW (superconducting) transition. The curves are vertically offset for clarity \citep{zhou2012evolution}. (b, c) $\sqrt{2}\times\sqrt{2}$ charge ordering ($V_{\mathrm{s}}$ = 40 mV, $I$ = 0.02 nA) and $dI/dV$ spectrum in KFe$_{2}$Se$_{2}$(001) films \citep{li2012KFeSe}.}
\end{figure}

Zhou \textit{et al} recently investigated the doping and temperature evolution of the electronic structure of 111-type NaFe$_{1-x}$Co$_{x}$As single crystals by STM spectroscopy [Fig.\ 6] \citep{zhou2012evolution}. In the parent compound NaFeAs, temperature-dependent $dI/dV$ spectra directly reveal a gap opening just below the spin density wave (SDW) transition temperature $T_{\mathrm{SDW}} \sim 40$ K. The observed gap edge peaks are consistent with the gap opening associated with a new electronic ordered state (such as superconductivity or SDW), where the DOS conservation will push states out toward the gap edge. In the case of the un-reconstructed surface of cleaved NaFeAs crystals, it is reasonably argued that the observed gap correlates closely with the SDW in the parent NaFeAs compound. Moreover, the SDW-related gap microscopically coexists and competes with the superconducting gap in an underdoped NaFe$_{1-x}$Co$_{x}$As ($x$=0.014) sample, which lies at the SDW and superconducting phase boundary \citep{cai2012visualizing}.

With increasing Co doping, the SDW gap is suppressed and an inhomogeneous superconducting gap occurs in optimally doped NaFe$_{1-x}$Co$_{x}$As ($x$=0.028). In contrast to cuprates, no pseudogap-like feature above $T_{c}$ is found in the $dI/dV$ spectra, similar to the doped FeSe$_{0.4}$Te$_{0.6}$(001) \citep{hanaguri2010unconventional} and stoichiometric FeSe(001), LiFeAs(001), and KFe$_{2}$Se$_{2}$(110) surfaces \citep{song2011direct,hanaguri2012scanning,chi2012scanning,li2011phase}. Unexpectedly, in overdoped NaFe$_{1-x}$Co$_{x}$As ($x$=0.061 and 0.075), a spatially inhomogeneous pseudogap-like feature re-enters and persists to the relatively high temperature of 50 K. Since no gap edge peaks can be clearly observed at positive bias voltages up to 50 meV, it is not sufficiently convincing to claim that the DOS depression around $E_{\mathrm{F}}$ originates from yet another ordered state. The nature of the DOS depression in overdoped NaFe$_{1-x}$Co$_{x}$As needs further investigation.

Using STM, Li \textit{et al} imaged four different phases in MBE-grown KFe$_{2}$Se$_{2}$(001) films on SrTiO$_{3}$ substrates. They argued that the parent compound corresponds to stoichiometric KFe$_{2}$Se$_{2}$ with $\sqrt{2}\times\sqrt{2}$ reconstruction with respect to the original Se lattice [Fig.\ 6(b)] \citep{li2012KFeSe}, although this surface had previously been suggested to be superconducting due to the observed V-shaped gap by Cai \textit{et al} \citep{cai2012imaging}. However, the nonvanishing DOS($E_{\mathrm{F}}$) at 0.4 K [Fig.\ 6(c)] suggests that the V-shaped gap of Cai \textit{et al} may have a non-superconducting origin \citep{li2012KFeSe}. Moreover, Cai \textit{et al} explained the $\sqrt{2}\times\sqrt{2}$ reconstruction as the consequence of the block antiferromagnetic state of the underlying Fe layer \citep{cai2012imaging}. Combining both studies, the V-shaped gap in the reconstructed $\sqrt{2}\times\sqrt{2}$ surface possibly originates from the SDW transition, similar to NaFeAs \citep{zhou2012evolution,cai2012visualizing}.

\section{\label{sec:QuasiparticleInterference}Quasiparticle interference}
In addition to inducing bound states, impurities can also scatter quasiparticles and lead to energy-dependent QPI patterns, which can be imaged in real space with spectroscopic STM \citep{fischer2007scanning,hoffman2011spectroscopic}. These patterns are then Fourier transformed to quantify the periodicity of electronic modulations. The combination of the real-space QPI pattern and momentum-space analysis is often called Fourier-transform scanning tunneling spectroscopy (FT-STS). Applications of FT-STS to cuprates have revealed dispersive or static electronic modulations in vortex cores, superconducting, and pseudogap states \citep{fischer2007scanning}. As a material enters into the superconducting state, its low-energy DOS redistributes according to the gap function $\Delta_{k}$, such that the QPI patterns at these energies contain detailed information about the orbital structure of the superconducting order parameter \cite{McElroy2003Relating}. Magnetic field-dependent FT-STS can even be used as a phase-sensitive probe to reveal the momentum-dependent coherence factors associated with quasiparticle scattering and thus distinguish between different pairing symmetries in unconventional superconductors \cite{hanaguri2010unconventional,hanaguri2009coherence}.

In Fe-SCs, FT-STS may present richer QPI patterns due to the multiple Fermi surfaces derived from the five Fe $d$-orbitals \citep{johnston2010puzzle}. STM on non-superconducting Ca(Fe$_{0.97}$Co$_{0.03}$)$_{2}$As$_{2}$ gives evidence for static, unidirectional electronic nanostructures aligned along the crystalline $a$ axis \citep{chuang2010nematic}, which are further suggested to arise from the combined effects of all electronic dimers induced by Co substitutions at Fe sites [Fig.\ 5(f)] \citep{allan2012anisotropic}. In fact, Song \textit{et al} more directly observed similar electronic dimers induced by Se substitutions at Fe sites in superconducting FeSe films [Figs.\ 5(d)] \citep{song2011direct,song2012suppression}. These findings provide evidence for the existence of a more complex electronic nematic state in the orthorhombic phase of some Fe-SCs, which may be responsible for the transport anisotropy in these materials \citep{allan2012anisotropic,chu2010plane}. Anisotropy in Fe-SCs without Fe-site dopants remains to be understood.

\begin{figure}[tbh]
\centering
\includegraphics[width=\columnwidth]{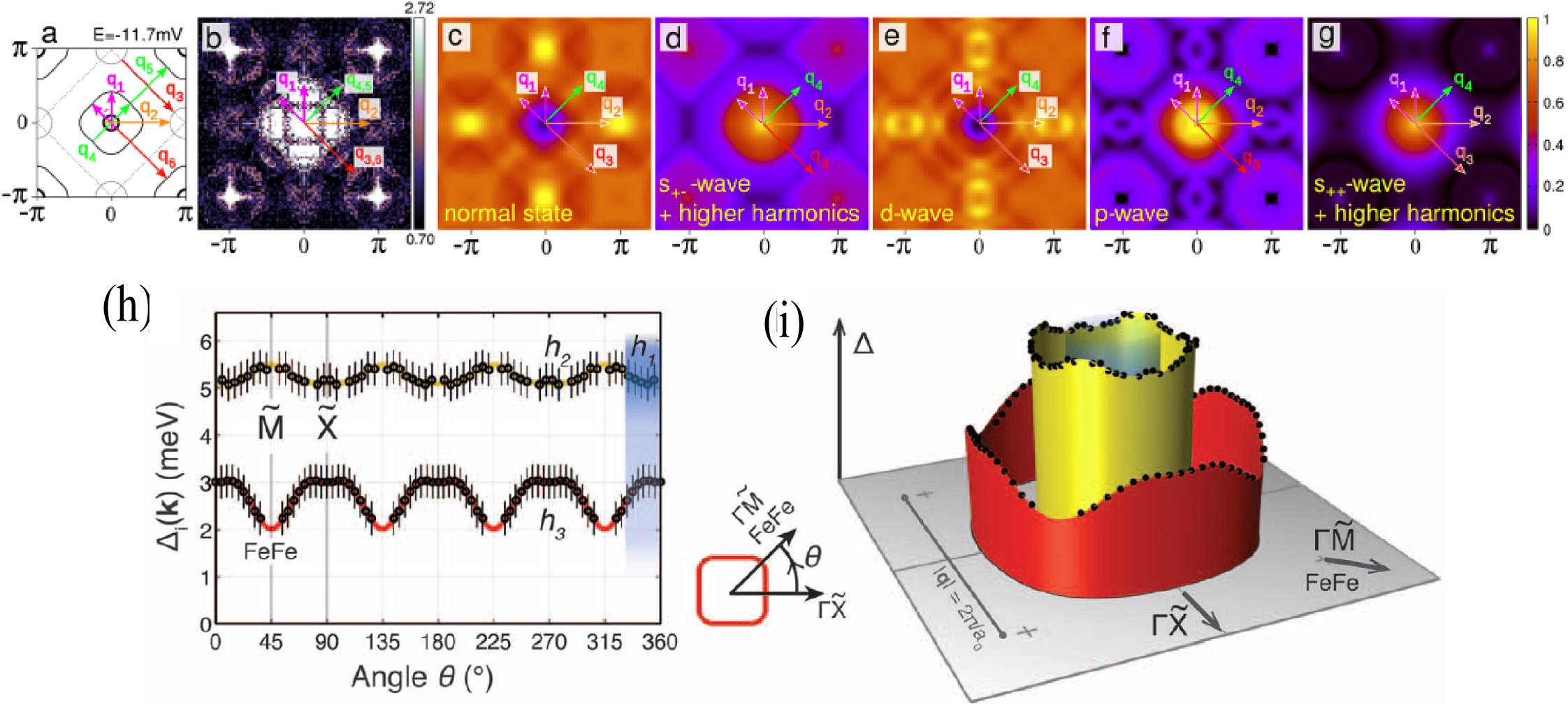}
\caption{(a) Schematic constant energy contours (CECs) of LiFeAs at -11.7 meV with the dashes indicating the first Brillouin zone. There are two hole-like pockets around $\Gamma$ and two electron-like pockets at the zone boundary. As marked, $q_{1...4}$ represent various scattering processes among the pockets, while $q_{5,6}$ represent umklapp processes. (b) Fourier transformed image of the measured QPI at the same energy as (a). The QPI features match well with the scattering $q_{1...6}$ vectors.
(c-g) Calculated QPI ($q$ space) in the normal and superconducting states with $s_{\pm}$, $d$, $p$, and $s_{++}$ wave pairing symmetries \citep{hanke2012probing}.
(h) Anisotropic energy gap structure $\Delta_{k}$ on the three hole-like bands $h_{1}$, $h_{2}$, and $h_{3}$ of LiFeAs single crystal at T = 1.2 K. The 0.35 meV error represents the thermal resolution of FT-STS at 1.2 K. (i) Three-dimensional image of the measured  $\Delta_{k}$ (solid dots) on the three hole-like bands \citep{allan2012science}. Four-fold symmetric superconducting gaps are clearly visible with their minima aligned along $\Gamma\widetilde{\mathrm{M}}$ ($\Gamma\widetilde{\mathrm{X}}$) for $h_{3}$ ($h_{1}$ and $h_{2}$) band.}
\end{figure}

Recently, two groups performed FT-STS studies of LiFeAs, chosen for its clean and charge neutral cleaved surface \citep{hanke2012probing,allan2012science}. H$\ddot{\mathrm{a}}$nke \textit{et al} first found that the Fourier transformed QPI images closely resemble the LiFeAs Fermi surface itself [Figs. 7(a) and 7(b)]. The QPI images are then explained as a series of interband scattering processes between a small $\Gamma$-centered pocket (which acts like a van Hove singularity) and other pockets \citep{hanke2012probing}. This contrasts with most other Fe-SCs where the scattering between the well-matched hole and electron pockets plays a crucial role in both QPI patterns and $s_{\pm}$-wave pairing \citep{hanaguri2010unconventional}.
To determine the superconducting order parameter from their experimental QPI images, H$\ddot{\mathrm{a}}$nke \textit{et al} calculated the expected QPI in the superconducting state with various elementary order parameters ($s_{++}$, $s_{\pm}$, $d$ and $p$ waves) within the BCS model. They suggested that their data supported $p$ wave pairing symmetry in LiFeAs, although a more complex order parameter such as $s+id$ wave could not be excluded. It should also be noted that the observation of a spin resonance in LiFeAs by neutron scattering does not support spin-triplet $p$-wave superconductivity in LiFeAs but instead suggests that the mechanism of superconductivity is similar to that in the other Fe-SCs \citep{taylor2011antiferromagnetic}.

Subsequently, Allan \textit{et al} measured intraband QPI in LiFeAs to characterize the anisotropic superconducting energy gap $\Delta_{k}$ \citep{allan2012science}. With much higher resolution than the previous experiment, they observed three distinct holelike $\Gamma$-centered bands $h_{1}$, $h_{2}$, and $h_{3}$. At specific energies, the Fourier transformed images of Bogoliubov QPI $g(q,E)$ exhibit disconnected slices with vanishing scattering intensity in certain directions for each band, evidencing anisotropic $\Delta_{k}$. Analysis shows that the gap minima correlate with the locus of maximum intensity in $g(q,E)$. Therefore, a plot of intensity maxima in a $g(q,E)$ plane along a high symmetry direction contains the information on $\Delta_{k}$. Figures 7(c) and 7(d) summarize the experimental results. The superconducting energy gap $\Delta_{k}$ for the $h_{3}$ ($h_{1}$ and $h_{2}$) band shows four-fold symmetry with its minima aligned along $\Gamma\widetilde{\mathrm{M}}$ ($\Gamma\widetilde{\mathrm{X}}$) orientation.  This differs from FeSe where two-fold pairing symmetry was suggested based on the anisotropic vortex structure \citep{song2011direct}.

\section{\label{sec:Vortexstate}Vortex State}
Vortices, quantized tubes of magnetic flux, form as a magnetic field penetrates into a type-II superconductor. Since the pioneering work by Hess \textit{et al} \citep{hess1989scanning}, STM has served as a powerful technique to visualize vortex arrangements in superconductors \citep{fischer2007scanning,yin2009scanning,hoffman2011spectroscopic}. More significantly, STM can uniquely access the internal structure of a single vortex core, providing a measure of the superconducting coherence length $\xi$ \citep{yin2009scanningprl,shan2011observation} and even fundamental information about electron pairing \citep{hanaguri2012scanning,song2011direct}.

In 1964, Caroli \textit{et al} theoretically predicted a series of quasiparticle bound states separated by $\sim\Delta^{2}/E_{\mathrm{F}}$ within the vortex cores \citep{caroli1964bound}. Since $E_{\mathrm{F}}$ is typically significantly larger than $\Delta$ in conventional superconductors \citep{shore1989density}, the excitation spectra are quasi-continuous, and will appear as a pronounced peak at $E_{\mathrm{F}}$, which can split into two symmetric peaks and eventually merge into the coherence peaks away from vortex center. This has been experimentally verified in NbSe$_{2}$ \citep{hess1989scanning,hess1990vortex}. In Fe-SCs, however, due to small carrier density (small $E_{\mathrm{F}}$) and high $T_{c}$ (large $\Delta$) \citep{johnston2010puzzle}, $\Delta^{2}/E_{\mathrm{F}}$ is relatively large and the excitation spectra are no longer continuous. Therefore, the quantum limit regime, where the DOS spectrum at the vortex center exhibit strong particle-hole asymmetry, may be anticipated \citep{shore1989density,hayashi1998low}.

\begin{figure}[tb]
\centering
\includegraphics[width=\columnwidth]{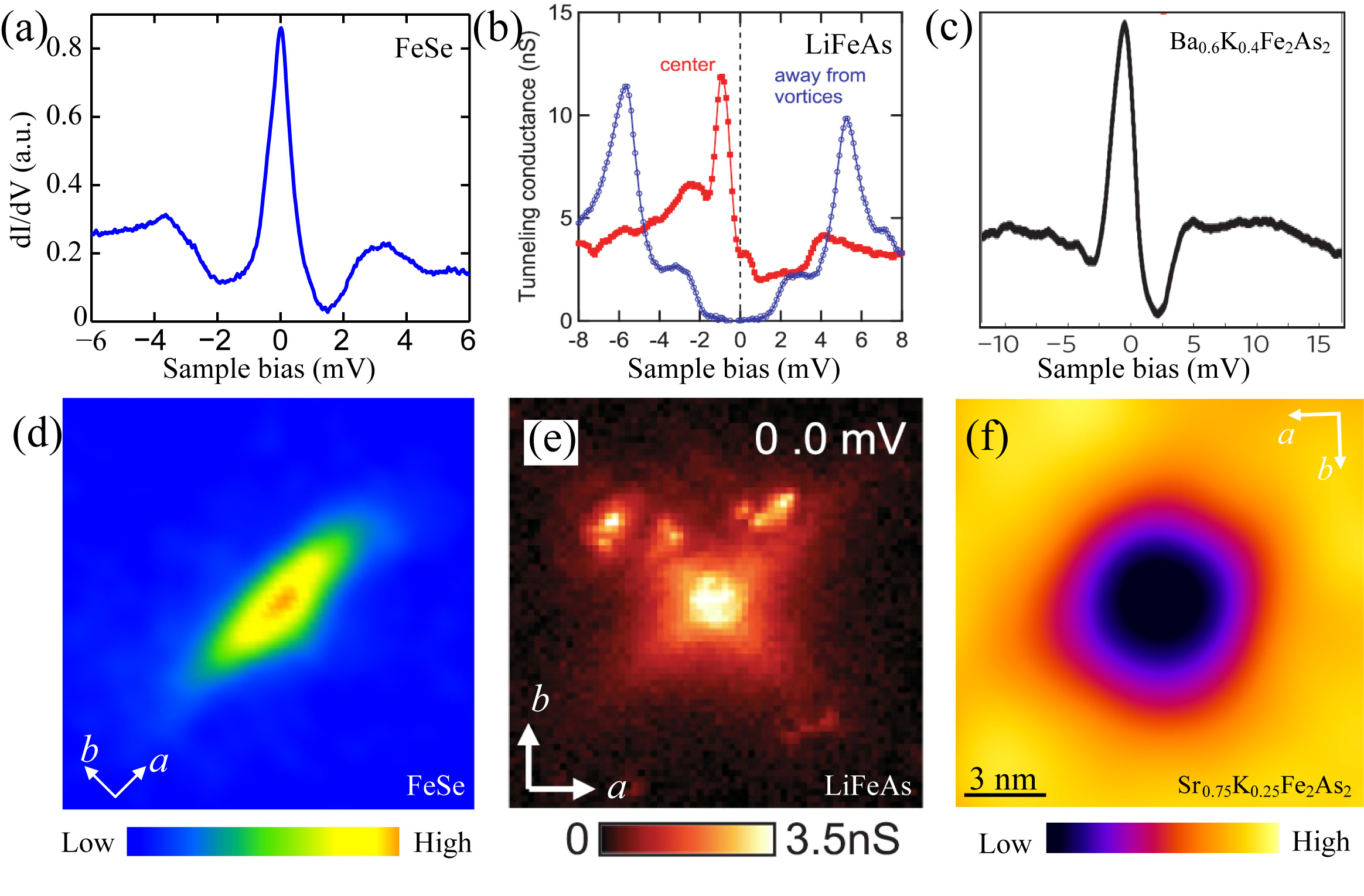}
\caption{ Vortex bound states and structure. (a-c) $dI/dV$ spectra taken at the vortex center in (a) FeSe \citep{song2011direct}, (b) LiFeAs \citep{hanaguri2012scanning}, (c) Ba$_{0.6}$K$_{0.4}$Fe$_{2}$As$_{2}$ \citep{shan2011observation}. (d-f) Vortex structure in (d) FeSe \citep{song2011direct}, (e) LiFeAs \citep{hanaguri2012scanning}, (f) Sr$_{0.75}$K$_{0.25}$Fe$_2$As$_2$ \citep{song2012electronic}. White arrows mark the nearest Fe-Fe directions.}
\end{figure}

STM has revealed bound states within the vortex cores of stoichiometric FeSe \citep{song2011direct} and LiFeAs \citep{hanaguri2012scanning}, as well as hole-doped Ba$_{0.6}$K$_{0.4}$Fe$_{2}$As$_{2}$ \citep{shan2011observation}. In FeSe, Song \textit{et al} observed a vortex core state exactly at $E_{\mathrm{F}}$ [Fig.\ 8(a)], which split into two symmetric peaks and eventually merged into the coherence peaks along the Fe-Fe orthorhombic $b$ axis, analogous to the conventional superconductor NbSe$_{2}$ \citep{hess1990vortex}. The splitting occurs along the Fe-Fe orthorhombic $a$ axis as well, but the two symmetric peaks never merge into the coherence peaks, providing evidence of the two-fold symmetry in electron pairing. In contrast, STM studies of LiFeAs and Ba$_{0.6}$K$_{0.4}$Fe$_{2}$As$_{2}$ revealed significant particle-hole asymmetry in the DOS spectrum of the bound states, with pronounced peaks below $E_{\mathrm{F}}$ [Figs.\ 8(b) and 8(c)]. This suggests that LiFeAs and Ba$_{0.6}$K$_{0.4}$Fe$_{2}$As$_{2}$ may be located in the quantum limit regime.

In addition, STM conductance maps reveal material-dependent vortex core shapes. In FeSe, the vortex appears elongated along the $a$-axis [Fig.\ 8(d)], leading to the suggestion that the electron pairing is two-fold symmetric in FeSe \citep{song2011direct}. In contrast, Hanaguri \textit{et al} identified a four-fold star-shaped vortex in LiFeAs, with high LDOS tails oriented at $45^{\circ}$ to the nearest Fe-Fe $a$ and $b$ axes [Fig.\ 8(e)] \citep{hanaguri2012scanning}. A model of gap anisotropy fails to account for the star-shaped vortex, because the gap minima in the outermost hole cylinder governing the LDOS distribution occur along the Fe-Fe directions \citep{allan2012science,umezawa2012unconventional} and thus should give rise to high LDOS tails along $a$ and $b$ axes, in stark contrast to the experiments. To address this confusion, Wang \textit{et al} theoretically calculated the shape of vortex cores by considering the Fermi surface anisotropy \citep{wang2012theory}. The results highlighted that the Fermi surface anisotropy can dominate the vortex core shape and prevent direct access of superconducting gap features if the gap itself is not highly anisotropic. Therefore, the observed four-fold star vortex core shape could be explained by the Fermi surface anisotropy in LiFeAs. Indeed, the ARPES experiment showed that the outermost hole cylinder exhibits a rounded square cross section with flat regions at $45^{\circ}$ to the Fe-Fe directions in LiFeAs \citep{umezawa2012unconventional}, which should lead to more quasiparticles and LDOS tails along these diagonals.

In nonstoichiometric Fe-SCs such as BaFe$_{1.8}$Co$_{0.2}$As$_{2}$ \citep{yin2009scanningprl}, Ba$_{0.6}$K$_{0.4}$Fe$_{2}$As$_{2}$ \citep{shan2011observation,wang2012close}, NaFe$_{0.975}$Co$_{0.025}$As \citep{wang2012close} and Sr$_{0.75}$K$_{0.25}$Fe$_2$As$_2$ \citep{song2012electronic}, single vortices often appear irregularly shaped and it is challenging to identify the vortex structure. To overcome this issue, Song \textit{et al} registered the vortex centers and then averaged the density of states from 48 vortices in Sr$_{0.75}$K$_{0.25}$Fe$_2$As$_2$ ($T_{c}$=32 K) \citep{song2012electronic}, as depicted in Fig.\ 8(f). Although the underdoped Sr$_{0.75}$K$_{0.25}$Fe$_2$As$_2$ lies in the orthorhombic phase \citep{johnston2010puzzle}, the averaged vortex appears nearly isotropic, in contrast to the two- and four-fold symmetric vortices in FeSe and LiFeAs \citep{hanaguri2012scanning,song2011direct}. This indeed supports the claim that the small orthorhombic lattice distortion can't bear the responsibility for the anisotropic vortices in FeSe \citep{song2011direct}. Further experiments are needed to understand the internal structure of single vortices in doped Fe-SCs and their bearing on the pairing symmetry.

\section{\label{sec:Conclusions}Conclusions}
We have shown that in the past year the STM has made significant contributions in the quest for fundamental understanding of the microscopic electron pairing mechanism in Fe-SCs. Low temperature tunneling spectroscopy has revealed: (i) a generic dip-hump spectral feature beyond the superconducting gap in all four main structural families of Fe-SCs \citep{song2011direct,hanaguri2010unconventional,fasano2010local,wang2012close,shan2012evidence,chi2012scanning}, which links to the spin resonance detected by neutron scattering \citep{wang2012close} and supports the importance of spin fluctuations in electron pairing in Fe-SCs; (ii) sub-gap bound states at single impurities in stoichiometric FeSe, LiFeAs and KFe$_{2}$Se$_{2}$ \citep{song2011direct,li2011phase,Grothe2012Bound}, consistent with a $s_{\pm}$ wave electron pairing picture; (iii) rotational symmetry breaking in both parent and superconducting states \citep{song2011direct,chuang2010nematic,zhou2011quasiparticle,allan2012anisotropic}, supporting an orbital ordering phase in Fe-SCs \citep{lv2009orbital,lee2009ferro}; (iv) SDW-related gaps opening near $E_{\mathrm{F}}$ in the Fe-SC parent compounds \citep{zhou2012evolution,cai2012visualizing,li2012KFeSe,cai2012imaging}. In addition to these achievements, QPI imaging has provided information about superconducting pairing symmetry \citep{hanaguri2010unconventional,hanke2012probing,allan2012science}, a prerequisite for the final determination of the nature of the high-$T_{c}$ superconductivity in Fe-SCs. Finally, STM has directly imaged material-dependent vortex core states and shapes \citep{shan2011observation,hanaguri2012scanning,song2011direct,song2012electronic}. Such results hint at the versatility in electron pairing, and can ultimately provide material-specific tests of the pairing symmetry in Fe-SCs.

Following these advances, there are a number of remaining questions to be addressed by the STM technique. In contrast to cuprates, the multi-band nature of Fe-SCs leads to electron pairing which relies sensitively on material and doping \citep{johnston2010puzzle}. Even within the same material, the superconducting order parameter can change from nodal, in thick FeSe films on graphene \cite{song2011direct}, to nodeless, in single unit-cell FeSe films on SrTiO$_{3}$ \citep{wang2012interface}. The grandest challenge remains to clarify the pairing symmetry and its origin, as a prerequisite to understanding the secret of high-$T_{c}$ superconductivity in Fe-SCs. Spectroscopy and QPI imaging serve as powerful techniques to extract the pairing symmetry. Meanwhile, spectroscopic imaging study of local DOS in the vicinity of both magnetic and non-magnetic impurities as well as vortices helps to distinguish between candidate pairing symmetries.

Another key issue is the role of dopants. For most Fe-SCs, particularly the higher-$T_c$ 1111- and 122-type materials, the superconductivity develops from the parent compounds upon chemical doping of electrons, holes, or even isovalent elements. In addition to enabling superconductivity, the dopant atoms are also potential sources of scattering, nanoscale phase separation, and electronic inhomogeneity, which may in turn lead to transport anisotropy, $T_{c}$ suppression, or vortex pinning. Therefore, STM can serve as an effective tool to probe these dopants on the nanometer scale and then understand their roles in superconductivity \citep{allan2012anisotropic,song2012electronic}.	

Finally, we anticipate that STM will play a crucial role in finding higher-$T_{c}$ materials. A number of tantalizing hints of higher-$T_{c}$ phases, such as 65 K $T_{c}$ in single layer FeSe on a SrTiO$_{3}$ substrate \citep{wang2012interface,he2012phase,tan2013interface} and 49 K $T_{c}$ in $\sim10\%$ volume fraction of Pr-doped CaFe$_{2}$As$_{2}$ \citep{lv2011unusual,saha2012structural}, have appeared in recent months. Just as STM served to disentangle the phase separation and identify the true chemical composition of the $T_{c}=32$ K superconducting state in KFe$_{2}$Se$_{2}$ \citep{li2011phase,li2012KFeSe}, STM should be employed to find the minority volume fraction superconductivity in rare-earth-doped CaFe$_{2}$As$_{2}$. Such attempts in both Ca$_{0.83}$La$_{0.17}$Fe$_{2}$As$_{2}$ \citep{huang2012experimental} and Ca$_{1-x}$Pr$_{x}$Fe$_{2}$As$_{2}$ (where no clear superconducting gap is observed) \cite{zeljkovic2013nanoscale} have so far been unsuccessful, but efforts are ongoing.

In conclusion, five years after the discovery of Fe-SCs, STM has contributed much to the understanding of these materials. Compared with the cuprates, Fe-SCs exhibit extraordinarily rich phenomenology, and can serve as a foil to unravel the secret of high-$T_{c}$ superconductivity. There remain several important open questions, to which STM has the unique potential to provide answers.

\section*{Acknowledgement}
%\begin{acknowledgments}
We thank J. C. Davis and H.-H. Wen for helpful conversations. C. L. Song was supported by the Golub Fellowship at Harvard University.
%\end{acknowledgments}



%% References
%%   \citep{key}          ==>>  [#]


\bibliographystyle{model3-num-names}
%\bibliography{FeSCsreview}

\end{document}